\documentclass[a4paper,showpacs,preprintnumbers,amsmath,amssymb,twocolumn]{revtex4}

\usepackage{graphicx}
\usepackage{dcolumn}
\usepackage{bm}
\usepackage{epstopdf}

\newcount\driver
\newcount\bozza

scaled\magstep1                  
scaled\magstep1 
 
scaled\magstep1                 \font\indbf=cmbx10 scaled\magstep2

scaled \magstep2

{\count255=\time\divide\count255 by 60
\xdef\hourmin{\number\count255}
        \multiply\count255 by-60\advance\count255 by\time
   \xdef\hourmin{\hourmin:\ifnum\count255<10 0\fi\the\count255}}

\let\a=\alpha \let\b=\beta    \let\g=\gamma     \let\d=\delta     \let\e=\varepsilon
  \let\h=\eta     \let\th=\vartheta \let\k=\kappa     \let\l=\lambda
\let\m=\mu    \let\n=\nu      \let\x=\xi                \let\r=\rho
\let\s=\sigma \let\t=\tau            
\let\ps=\psi   \let\o=\omega     
 \let\D=\Delta       \let\L=\Lambda    \let\X=\Xi

\def\EE{{\cal E}}\def\VV{{\cal V}}
\def\WW{{\cal W}}
\def\BB{{\cal B}}
\def\RR{{\cal R}}\def\LL{{\cal L}}

\def\pp{{\bf p}}\def\xx{{\bf x}}
   
\def\yy{{\bf y}}\def\kk{{\bf k}}\def\nn{{\bf n}}
\def\zz{{\bf z}}

       \def\oo{{\underline \omega}}
\def\ee{{\underline \varepsilon}}

        \def\EE{\hbox{\msytw E}}

\let\==\equiv

\let\io=\infty
\let\0=\noindent

\def\*{{\hfill\break\null\hfill\break}}

\def\media#1{{\langle#1\rangle}}

\def\tilde#1{{\widetilde #1}}

\def\tende#1{\,\vtop{\ialign{##\crcr\rightarrowfill\crcr
             \noalign{\kern-1pt\nointerlineskip}
             \hskip3.pt${\scriptstyle #1}$\hskip3.pt\crcr}}\,}
\def\otto{\,{\kern-1.truept\leftarrow\kern-5.truept\to\kern-1.truept}\,}

\def\Tr{\rm Tr}

\def\wh#1{\widehat{#1}}
\def\hat#1{\wh{#1}}
\def\sqt[#1]#2{\root #1\of {#2}}

\def\bp{{\bar \ps}}

\def\EE{{\cal E}}\def\VV{{\cal V}}
\def\WW{{\cal W}}
\def\BB{{\cal B}}
\def\RR{{\cal R}}\def\LL{{\cal L}}

\def\T#1{{#1_{\kern-3pt\lower7pt\hbox{$\widetilde{}$}}\kern3pt}}
\def\VVV#1{{\underline #1}_{\kern-3pt
\lower7pt\hbox{$\widetilde{}$}}\kern3pt\,}
\def\W#1{#1_{\kern-3pt\lower7.5pt\hbox{$\widetilde{}$}}\kern2pt\,}

\def\indica{\leaders \hbox to 0.5cm{\hss.\hss}\hfill}
\def\guida{\leaders\hbox to 1em{\hss.\hss}\hfill}
\mathchardef\oo= "0521

\def\pp{{\bf p}}\def\xx{{\bf x}}
\def\yy{{\bf y}}\def\kk{{\bf k}}\def\nn{{\bf n}}
\def\zz{{\bf z}}

\def\oo{{\underline \omega}}

\def\qed{\raise1pt\hbox{\vrule height5pt width5pt depth0pt}}
 
  \def\bp{{\bar p}} 

\def\indic{\hbox{\raise-2pt \hbox{\indbf 1}}}

\def\ins#1#2#3{\vbox to0pt{\kern-#2 \hbox{\kern#1 #3}\vss}\nointerlineskip}

\newdimen\xshift \newdimen\xwidth \newdimen\yshift

\def\insertplot#1#2#3#4#5#6{%
\xwidth=#1pt \xshift=\hsize \advance\xshift by-\xwidth \divide\xshift by 2%
\begin{figure}[ht]
\vspace{#1pt} \hspace{\xshift}
\\begin{center}
\begin{minipage}{#1pt}
#3 \ifnum\driver=1 \griglia=#6
\ifnum\griglia=1 \openout13=griglia.ps \write13{gsave .2
setlinewidth} \write13{0 10 #1 {dup 0 moveto #2 lineto } for}
\write13{0 10 #2 {dup 0 exch moveto #1 exch lineto } for}
\write13{stroke} \write13{.5 setlinewidth} \write13{0 50 #1 {dup 0
moveto #2 lineto } for} \write13{0 50 #2 {dup 0 exch moveto #1
exch lineto } for} \write13{stroke grestore} \closeout13
\includegraphics{griglia.ps} \fi
\includegraphics{#4.ps}\fi%
\ifnum\driver=2 \fi
\end{minipage}
r
\end{center}
\end{figure}
}

\newdimen\shift \shift=-1.5truecm
\def\lb#1{%
\ifnum\bozza=1
\label{#1}\rlap{\hbox{\hskip\shift$\scriptstyle#1$}}
\else\label{#1} \fi}

\def\be{\begin{equation}}
\def\ee{\end{equation}}
\def\bea{\begin{eqnarray}}\def\eea{\end{eqnarray}}
\def\bean{\begin{eqnarray*}}\def\eean{\end{eqnarray*}}
\def\bfr{\begin{flushright}}\def\efr{\end{flushright}}
\def\bc{\begin{center}}\def\ec{\end{center}}
\def\bal{\begin{align}}\def\eal{\end{align}}
\def\ba#1{\begin{array}{#1}} \def\ea{\end{array}}
\def\bd{\begin{description}}\def\ed{\end{description}}
\def\bv{\begin{verbatim}}\def\ev{\end{verbatim}}
\def\nn{\nonumber}
\def\Halmos{\hfill\vrule height10pt width4pt depth2pt \par\hbox to \hsize{}}
\def\pref#1{(\ref{#1})}

\def\ins#1#2#3{\vbox to0pt{\kern-#2 \hbox{\kern#1 #3}\vss}\nointerlineskip}

\newdimen\xshift \newdimen\xwidth \newdimen\yshift
\newcount\griglia

\def\insertplot#1#2#3#4#5#6{%
\xwidth=#1pt \xshift=\hsize \advance\xshift by-\xwidth \divide\xshift by 2%
\begin{figure}[ht]
\vspace{#2pt} \hspace{\xshift}
\begin{minipage}{#1pt}
#3 \ifnum\driver=1 \griglia=#6
\ifnum\griglia=1 \openout13=griglia.ps \write13{gsave .2
setlinewidth} \write13{0 10 #1 {dup 0 moveto #2 lineto } for}
\write13{0 10 #2 {dup 0 exch moveto #1 exch lineto } for}
\write13{stroke} \write13{.5 setlinewidth} \write13{0 50 #1 {dup 0
moveto #2 lineto } for} \write13{0 50 #2 {dup 0 exch moveto #1
exch lineto } for} \write13{stroke grestore} \closeout13
\includegraphics{griglia.ps} \fi
\includegraphics{#4.ps}\fi%
\ifnum\driver=2 \fi
\end{minipage}
\caption{#5}
\end{figure}
}

\newdimen\shift \shift=-1.5truecm
\def\lb#1{%
\label{#1}\rlap{\hbox{\hskip\shift$\scriptstyle#1$}}
\else\label{#1} \fi}

\def\be{\begin{equation}}
\def\ee{\end{equation}}
\def\bea{\begin{eqnarray}}\def\eea{\end{eqnarray}}
\def\bean{\begin{eqnarray*}}\def\eean{\end{eqnarray*}}
\def\bfr{\begin{flushright}}\def\efr{\end{flushright}}
\def\bc{\begin{center}}\def\ec{\end{center}}
\def\bal{\begin{align}}\def\eal{\end{align}}
\def\ba#1{\begin{array}{#1}} \def\ea{\end{array}}
\def\bd{\begin{description}}\def\ed{\end{description}}
\def\bv{\begin{verbatim}}\def\ev{\end{verbatim}}
\def\nn{\nonumber}
\def\Halmos{\hfill\vrule height10pt width4pt depth2pt \par\hbox to \hsize{}}
\def\pref#1{(\ref{#1})}

\usepackage{amsmath}
\usepackage{amsfonts}
\usepackage{amssymb}
\usepackage{epstopdf}

scaled\magstep1

\let\a=\alpha \let\b=\beta  \let\g=\gamma  \let\d=\delta
\let\e=\varepsilon
  \let\h=\eta   \let\th=\theta \let\k=\kappa \let\l=\lambda
\let\m=\mu    \let\n=\nu    \let\x=\xi         \let\r=\rho
\let\s=\sigma \let\t=\tau    
\let\ps=\Psi   \let\o=\omega
 \let\D=\Delta  \let\L=\Lambda \let\X=\Xi

\def\EE{{\cal E}} \def\VV{{\cal V}}
 \def\WW{{\cal W}}
 
\def\RR{{\cal R}}\def\LL{{\cal L}}

 \def\pp{{\bf p}}
 \def\xx{{\bf x}} \def\yy{{\bf y}} \def\zz{{\bf z}}
\def\kk{{\bf k}}

\def\nn{\nonumber}

\def\\{\hfill\break}
\def\={:=}
\let\io=\infty
\let\0=\noindent
\def\media#1{{\langle#1\rangle}}

\def\tende#1{\,\vtop{\ialign{##\crcr\rightarrowfill\crcr\noalign{\kern-1pt
    \nointerlineskip} \hskip3.pt${\scriptstyle #1}$\hskip3.pt\crcr}}\,}
\def\otto{\,{\kern-1.truept\leftarrow\kern-5.truept\to\kern-1.truept}\,}

\def\wh{\widehat}
\def\to{\rightarrow}

\def\qed{\hfill\raise1pt\hbox{\vrule height5pt width5pt depth0pt}}

\def\be{\begin{equation}}
\def\ee{\end{equation}}
\def\bp{\begin{pmatrix}}
\def\ep{\end{pmatrix}}
\def\bea{\begin{eqnarray}}
\def\eea{\end{eqnarray}}
\def\nn{\nonumber}
\def\pref#1{(\ref{#1})}

\def\lb{\label}

\def\Tr{\mathrm{Tr}}

\begin{document}

\title{Persistence of gaps in the interacting Hofstadter model}
\author{V. Mastropietro}
\affiliation{Universit\`a di Milano, Via Saldini 50, 
Milano, Italy}

\begin{abstract} 
The energy spectrum of the Hofstadter model has a fractal structure with
infinitely many gaps. 
We prove the persistence of each gap in presence of Hubbard interaction in the case of small transversal hopping, even when the coupling 
is 
much larger than the non interacting gaps.
The proof relies on a subtle interplay of
Renormalization Group arguments combined
with number-theoretic properties of the incommensurate frequencies.
\end{abstract}


\maketitle


\section{Introduction}

The energy spectrum
of non-interacting 
electrons in two dimensions moving through periodic magnetic and 
electrostatic potentials provides one of few example of fractals in quantum physics. A paradigmatic example is provided by the
{\it Hofstadter model} \cite{h},  \cite{T}
describing fermions hopping on a square lattice with a magnetic field in the orthogonal direction. 
The crucial parameter is the ratio $\a$ between the magnetic and electrostatic
lengths. If $\a$ is rational the
two lengths are commensurate and Bloch theory predicts a finite number of gaps.
In correspondence of such gaps one has an integer Hall conductivity \cite{T}.
If one considers sequences of rationals $\a$ converging to an irrational,
more and more gaps open and at the end 
a fractal spectrum with a self-similar
structure appears. Such features are found in absence of interaction between particles, which should be taken into account in any realistic description.
Indeed recent experimental realizations of 
the Hofstadter model
in graphene \cite{1}, \cite{2}, \cite{3} or 
photon systems \cite{4} exhibit important signatures of many body effects.

Interactions in the Hofstadter model
in the {\it commensurate} case have been analyzed by mean field or lowest order analysis 
in \cite{6}-\cite{10}. In \cite{KM}-\cite{TK1} the anisotropic Hofstadter model was effectively described in terms of 
an array of wires, and the continuum limit,
where the difference between the commensurate or incommensurate case is lost, makes possible a bosonization approach.
Incommensurability effects are however known
to be crucial in the Hofstadter model; its properties 
can be deduced by the one dimensional {\it  Aubry-Andre' } \cite{AA} model or the (equivalent) {\it  Harper} 
or {\it  almost-Mathieu} equation:
when $\a$ is irrational its spectrum is a 
Cantor set and a
delocalization-localization transition is present, see e.g. \cite{DS}-\cite{A}. Such features
are connected with the similarity of the properties of the almost-Mathieu equation with the 
{\it Kolmogorov-Arnold-Moser} (KAM) theorem, providing another celebrated example of fractal structures in physics. 
The effect of the interaction in generalized
Aubry-Andre' models  in the extended regime
has been studied in \cite{M1}, and in the related case of interacting fermions with a Fibonacci potential in \cite{G1} \cite{G2}; the interacting Aubry-Andre' model
in the localized regime has been considered
in \cite{Hu},\cite{M2} and in a dynamical context in
\cite{z3}-\cite{z1}.
The 
equivalence between Hofstadter and Aubry-Andre' model is however lost in presence of a many body interaction, which makes the problem not reducible to a one dimensional system.

A natural question is if the 
fractal structure of the spectrum of the Hofstadter mode
is preserved by the many body interaction, or is instead an artifact of the single particle approximation.
This is equivalent to ask if 
the gaps present in the non interacting case are or not closed by the interaction. 
If the interaction is larger than the gaps this cannot happen, but in the present case all non interacting gaps except a finite number are smaller than any finite interaction.
In the related one dimensional Fibonacci chain,
a scenario was indeed suggested in  \cite{G1}, \cite{G2} according to which all  gaps except a finite number are closed by any attractive interactions
and the spectrum acquire finite measure.
The question of persistence of gaps
is related to a
{\it small divisor problem}, caused by processes involving large exchange of momentum such that, due to Umklapp,
connect with arbitrary precision the Fermi points. Small divisors appears in
KAM theory and make the problem  
non-perturbative; physical properties cannot be understood by lowest order analysis but are encoded in the divergence or convergence of the whole perturbative series. Typical examples of small divisor problems in classical mechanics are the Birkhoff series for prime integrals 
of perturbed integrable Hamiltonian system, which are typically diverging (Poincare' theorem), or the series for KAM tori which are instead convergent \cite{Ga}.

In order to get information on the spectrum we compute
the large distance behavior of the 2-point function
in the interacting Hofstadter model for values of the chemical potential corresponding to the gaps of the non-interacting case.
The persistence of the gaps is detected by the presence of a faster
than any power large distance decay. This approach allows us 
to use non-perturbative Renormalization Group (RG) methods.
The main difficulty relies in the fact that 
the incommensurability produces an infinite set of effective interactions almost connecting the Fermi points, and the persistence or not of gaps is connected by their relevance or irrelevance in the RG sense. 

The rest of this paper is organized as follows.
In \S 2 we introduce the model and we present the main result. In \S 3 we
recall the main features of the non interacting case.
In \S 4 we analyze the Euclidean correlations of the interacting model
by rigorous Renormalization Group methods. In \S 5 we show the convergence of the RG iteration, using a Diophantine property for $\a$. Finally in \S 6 the 
main conclusions are presented.

\section{The  interacting Hofstadter model}

We consider an interacting version of the Hofstadter model in which 
spinful fermions in a square lattice with step one are subject to a magnetic field $\vec A=(-B x_2,0,0)$  and interact through 
a Hubbard interaction.

The Hamiltonian of the (anisotropic) Hofstadter-Hubbard model is
$H=H_0+V$ with $H_0=$
\bea
&&-{1\over 2}\sum_{\vec x,\s=\uparrow, \downarrow} 
(a^+_{\vec x+ \vec e_1,\s} e^{-i 2\pi \a x_2}
 a^-_{\vec x,\s}+a^+_{\vec x,\s} e^{i 2\pi \a x_2}
 a^-_{\vec x +\vec e_1,\s})-\nn\\
&&\sum_{\vec x,\s=\uparrow, \downarrow}
(t (a^+_{\vec x+\vec e_2,\s}
a^-_{\vec x,\s}+a^+_{\vec x,\s}
a^-_{\vec x+\vec e_2,\s})-\m 
a^+_{\vec x,\s}
a^-_{\vec x,\s})
\eea
where $a^\pm_{\vec x,\s}$ are fermionic operators, $\s$ is the spin,
$\vec x=(x_1,x_2)$ are points in a square lattice with step $1$ (
pbc in the $1$ direction and Direchelet in direction $2$), $t>0$ is the hopping in the 2-direction,
$\m=\cos p_F$ is the chemical potential,
$B=2\pi \a$ is the vector potential and the interaction is
\be
V=U \sum_{\vec x}  a^+_{\vec x,\uparrow} a^-_{\vec x,\uparrow}a^+_{\vec x,\downarrow} 
a^-_{\vec x,\downarrow} 
\ee
%
with $U\ge 0$. In the $t=0$, the multi-wire limit, the system reduces to uncoupled one dimensional interacting chains parametrized by $x_2$, with different chemical potential.
If $t=1/2$ then $H_0$
is the Hofstadter Hamiltonian 
\cite{h}, and $t\not=1/2 $ case corresponds to the anisotropic generalization 
in \cite{T}. The Hamiltonian can be written as $H_0=\sum_{k_1}
H_0(k_1)$; the eigenfunctions of $H_0(k_1)$ are Slater determinants of the eigenfunctions of the (single-particle) one dimensional almost-Mathieu 
or Aubry-Andre' equation, parametrized by $k_1$
\be
-t(u_{x_2-1}+u_{x_2+1})-\cos (k_1-2\pi \a x_2) u_{x_2}=E u_{x_2}
\ee
The properties of  
the Aubry-Andre' equation depend crucially whether $\a$ is rational or irrational: in the second case 
the spectrum is a Cantor set \cite{DS}, \cite{FSW}, \cite{A} 
and has pure point spectrum with exponentially decaying eigenfunctions 
for $t<1/2$, purely singular-continuous spectrum for $t=1/2$, 
purely absolutely continuous spectrum for $t>1/2$.  The gaps of $H_0$  are located in correspondence of \cite{T} 
\be
p_F=n_F \pi \a\quad {\rm mod}\quad 2\pi \label{aa}
\ee
with $n_F$ integer; equivalently 
\pref{aa} can be written as 
$N/N_0=n_F \a+s$, with $N_0$ the maximal number of fermions and $s$ integer. 

When the interaction is present $U\not=0$ the system is not reducible 
to a one dimensional one. Information on the spectrum can be obtained by
the large distance decay of imaginary time correlations.
If  $a^\pm_{\bar x,\s}=e^{H x_0} a^\pm_{\vec x,\s} e^{-H x_0}$ with $\underline x=(\xx,x_2)$, $\xx=x_0,x_1$,
the zero temperature 2-point is 
$S(\underline x,\underline y)=< a^-_{\underline x,\s}
a^+_{\underline y,\s}>$ with $<O>=\lim_{\b\to\io,L\to\io} Tr e^{-\b H} T O/
Tr e^{-\b H}$, $T$ is time ordering. 


It is convenient to choose $\a$ as 
{\it Diophantine}, that is there exists $C_0,\t$ such that
\be ||2 n \pi\a||\ge C_0
|n|^{-\t},\quad\quad n\not=0\label{d}\ee 
$||.||$ being the norm on the one dimensional $2\pi$ torus.
Any irrational except a zero measure set verifies such a property for
some $C_0,\t$. This kind of condition is usually assumed to deal
with small divisors, like in KAM or in the almost Mathieu equation.
We assume that $t$ is small, that is we are close to the multi-wire limit, corresponding to the localized regime in the non interacting case.

Our main result is the following
\vskip.3cm
{\it Chosen $p_F$ verifying \pref{aa} and $\a$ verifying
\pref{d}, 
for $t, U$ small and positive
$S(\underline x,\underline y)$ decays for large 
distances as, for any $N$
\be
|S(\underline x,\underline y)|\le {1\over |\xx-\yy|^{1+\h}}{C_N\over 
1+(\bar \D|x_2-y_2|+\D|\xx-\yy|)^N}
\ee
with $\bar\D=|\log t|$ and
\be
\D=
t^{n_F}(a_{n_F}+R)\label{gg}
\ee
with $|R|\le  C(t+U^2)$, $a_{n_F}$ non vanishing and independent on $x_2$.
}
\vskip.3cm
The faster than any power decay in the imaginary time 
signals the presence of a gap in the spectrum
of the interacting Hofstadter model; the decay rate $\D$ provides an estimate of the gap.
For small $t$ (smaller than $a_{n_F}$), that is close to the multi-wire limit, the gaps in the non interacting case are
proportional to a power of the hopping, that is are $O(a_{n_F}t^{n_F})$. Our result shows that 
the gaps {\it persist} even in presence of an interaction
{\it much larger} than the non interacting gap, provided that $U$ is smaller than $a_{n_F}$, that is $a_{n_F}t^{n_F}<<U^2<<a_{n_F}$.

The decay in the $2$ direction is much faster, and is in agreement with the fact that
in absence of interaction the system is in the localized regime, with eigenfunctions decaying with rate $\log t$. In the direction $1$  at distances smaller than the inverse of the gap there is a power 
law decay with an anomalous exponent $\h$, signaling that in the decoupled limit $t=0$ the system reduces to arrays of interacting fermionic wires. 

Note that a critical exponent appear in the decay in the $1$ direction, as consequence of the Luttinger liquid behavior of the multi-wire limit, but not
in the gaps, as it would be in gapped one dimensional Luttinger liquids
and happens in the interacting Aubry-Andre' model \cite{M1}, as a consequence
of the planarity of the Hofstadter model. As in \cite{KM}-\cite{TK1}, we are considering the anisotropic case close to multi-wire limit
$t=0$, where the system decouples in independent wires; in contrast to  
\cite{KM}-\cite{TK1}, however, we do not perform any continuum limit but we are
taking into full account lattice effects, which are essential to distinguish between the commensurate and incommensurate case.

\section{Small divisors and Feynman graphs} 

The persistence of gaps is studied expanding the imaginary-time correlations around the point $U=t=0$, where the system reduces to a collection of independent fermionic wires labeled by $x_2$ 
with dispersion relation 
$\cos (k_1-2\pi \a x_2 )$; the Fermi points are given by 
\be
p_\pm^{x_2}=\pm p_F+2\pi\a  x_2
\ee
if $\m=\cos p_F$. The 2-point function 
$S(\underline x,\underline y)|_{t=U=0}\equiv \bar g(\underline x,\underline y)$
is
\be
\bar g(\underline x,\underline y)
=\d_{x_2,y_2}
\int d\kk e^{i\kk(\xx-\yy)} \hat g_{x_2} (\kk)\label{jj1}
\ee
where
\be 
\hat g_{x_2} (\kk)={1
\over -ik_0+\cos (k_1-2\pi\a x_2 )-\cos  p_F }\label{pp}
\ee
It is convenient to write the imaginary-time correlations in terms of the following Grassmann integral
\be
e^{W(\phi)}=\int P(d\psi) e^{-T-V-N-(\psi,\phi)}
\ee
with
\bea
&&T=\sum_{x_2,\s} \int d\xx (\psi^+_{\xx,x_2+1,\s} \psi^-_{\xx,x_2,\s}+ 
\psi^+_{\xx,x_2-1,\s} \psi^-_{\xx,x_2,\s})\nn\\
&&V=U \sum_{x_2,\s}\int d\xx 
\psi^+_{\xx,x_2,\uparrow} \psi^-_{\xx,x_2,\uparrow}\psi^+_{\xx,x_2,\downarrow} 
\psi^-_{\xx ,x_2,\downarrow}\\
&&N=\sum_{x_2} \n_{x_2}  \int d\xx (\psi^+_{\xx,x_2,\s} \psi^-_{\xx,x_2,\s}+ 
\psi^+_{\xx,x_2,\s} \psi^-_{\xx,x_2,\s})\nn 
\eea
and $(\psi,\phi)=\sum_{x_2}\int d\xx (\psi^+_ {\xx,x_2,\s}\psi^-_ {\xx,x_2,\s}+\psi^-_ {\xx,x_2,\s}\psi^+_ {\xx,x_2,\s}$. The term $N$ has been introduced
writing the chemical potential as $\m=\cos p_F+\n$, in order to take into account its possible renormalization due to the interaction.
The 2-point function is given by 
$S(\underline x,\underline y)= {\partial^2 W\over \partial  \phi^+_{\underline x}\partial  \phi^-_{\underline x} }|_0$.
\insertplot{700}{180}
{}%
{figjsp467aa}
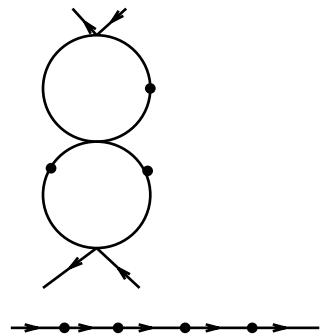
{\label{n9} A graph with four external lines of order $t^3 U^3$ and another with two external lines of order $t^4$.
}{0}
One can write the correlation in terms of Feynman diagrams with propagators \pref{jj1} ; examples are in Fig. 1. The small divisors problem is clearly exhibited 
already in the non-interacting case 
$U=0$. Consider a chain graph contributing to the effective potential 
$\int d\kk \phi^+_{x_2,\kk} W_2(\kk) \phi^-_{x'_2,\kk}$ with $x'_2=x_2+\sum_{k=1}^{n}\e_k$ and
$W_2(\kk)=$
\be
[t^n  
\prod_{k=1}^{n-1} {1
\over -ik_0+\cos (k-2\pi\a (x_2+\e_k) )-\cos  p_F }]\label{zak}
\ee
The infrared divergences in many body perturbation theory are associated with the repetitions of propagators with the same
momentum $k'$ measured from the Fermi points, that
is $k_1=k'+p_\pm^{x_2}$; if $x_2$ and $x'_2$ are the coordinates associated to two propagators, this happens if $x_2=x'_2$, $\o=\o'$, $\o=\pm$
or, if $p_F=n_F\pi \a$, if $x_2-x'_2=-\o n_F$ and $\o=-\o'$:
in such cases the subgraph are resummed in the self energy or the mass terms.
If $\a$ is rational, if  $x_2-x'_2\not=0, \o n_F$ the denominators
differ by a finite quantity $O(1/q)$ if $\a=p/q$ with $p,q$ coprime. 
If $\a$ is irrational, however, $2\pi\a(x_2-x'_2)$ can be  
{\it arbitrarily close} mod. $2\pi$ to $0$ or $2 n_F\pi \a$; in other words, due to Umklapp terms involving the exchange of $2\pi$,
there are propagators with almost the same size which cannot be resummed in self energy or mass terms. This produces an accumulation of small divisors which could cause a failure of the expansion.

Consider for instance the case $\e_k=1$ in \pref{zak}; then the momenta flowing in the propagators would be all different. 
By the diophantine condition, if $k=n_F\pi\a$ we can bound each propagator by  
\be
|\hat g_{x_2} (\kk)|\le {C\over ||2\pi \a x_2+2\pi \a n_F||}\le C |x_2+n_F|^\t
\ee
so that 
\be
|W_2(\kk)|\le C^n t^n \prod_{k=1}^n k^\t\le  C^n t^n n!^\t
\ee
The appearance of such factorials, possibly breaking the convergence of the series,
is what is known in classical mechanics as small divisors. Physical information cannot be decided on the basis of lowest order analysis, but it depends on the convergence or divergence of the whole series.
Formal series for prime integrals in perturbed 
integrable Hamiltonian systems  are order by order finite
but typically non convergent, that is no prime integrals except the energy exists (Poincare' theorem). In other cases, instead, 
the bounds can be improved and the factorials cancel out; this is what happens in Lindstedt series for KAM tori. This is also what happens in
the Hofstadter model, where convergence of perturbation theory
is implied by results on the almost Mathieu equation using KAM methods. The persistence of the gap in the interacting Hofstadter model
depends on the convergence or divergence of its series expansions, which contains also graphs with loops in addition to chain graphs, and cannot be decided on the basis of lowest order perturbative considerations.


\section{Renormalization Group analysis} 

We study the 2-point function of the interacting Hofstatder model by exact RG methods. The starting point is the multiscale decomposition of the propagator
\be
g_{x_2} (\xx,\xx')=g^{(1)}_{x_2} (\xx,\xx')+\sum_{\o=\pm}
g^{(\le 0)}_{\o; x_2} (\xx,\xx') 
\ee
where $\hat g^{(\le 0)}_{\o; x_2} (\kk)$ has support in a region around $(0,p_\o^{x_2})$, $\o=\pm$,
and $\hat g^{(1)}_{x_2} (\kk)$ in the complement of such regions.

It is convenient to measure the momenta from the Fermi points
writing $k_1=k'+\o p_F+2\pi\a x_2$; therefore 
$\psi=\psi^1+\sum_{\o=\pm} e^{i p_\o^{x_2}x_1}\bar \psi_\o^{(\le 0)}$
and the propagator of $\bar \psi_\o^{(\le 0)}$ is 
\be
g^{(\le 0)}_{\o} (\underline x,\underline x') =\d_{x_2,x'_2}\int d\kk' 
{  \chi_0(\kk') e^{i\kk'(\xx-\xx')}\over -ik_0\pm v_F k'
+r(k')}
\ee	
with $r(k')=O(k'^2)$ and $\chi(\kk')$ has support around $\kk'=0$.

Integrating the scales $\le 0$ we get a sequence
of effective potentials sum of terms of the form $\int W_n \prod_{i=1}^n \psi^{\e_i}_{\o_i,\kk'_i,x_{2,i} }$, $\e,\o=\pm$, with momenta $\kk'_i$ verifying the relations
\be
\sum_i \e_i k'_i=
\sum_i \e_i \o_i p_F+\sum_i \e_i  2\pi \a x_{2,i}\quad {\rm mod.} 2\pi\label{delta}
\ee
Note that the momenta measured from the Fermi points are not conserved unless the r.h.s. of the above expression is vanishing.

Note that after the integration of $\psi^0$ 
a mass term, which was absent in the original interaction, is generated, of the form
\be
\sum_{x_2}  \int d\kk' W_2(\kk')
(\psi^{(\le 0), +}_{+,x_2-n_F,\kk'}\psi^{(\le 0), -}_{-,x_2,\kk'}
+\psi^{(\le 0), +}_{-,x_2,\kk'}\psi^{(\le 0), -}_{+,x_2-n_F,\kk'})\ee
which connect fields in chains $x_2,x_2-n_F$, with momenta near
$p_{x_2-n_F}^+=p_F+2\pi \a (x_2-n_F)$
to $p_{x_2}^-=-p_F+2\pi \a (x_2)=p_{x_2-n_F}^+$. The lowest order contribution is the chain graph $G_{x_2-n_F,x_2} (0,p_{x_2-n_F}^+)$
, see Fig.2, with 
\be
G^0_{x_2-n_F,x_2} (\kk)=
t^{n_F}  g_{x_2-n_F+1}^{(0)}(\kk) g_{x_2-n_F+2}^{(0)}(\kk)...g_{x_2-1}^{(0)}(\kk)
\ee
\insertplot{500}{120}
{
\ins{20pt}{90pt}{$x_2$}
\ins{90pt}{90pt}{$x_2-n_F$}
}%
{figjsp467bb}
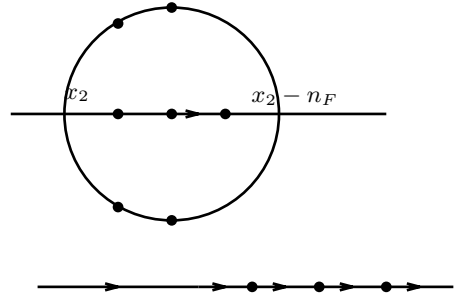
{\label{n9} The upper graph is a contribution to the mass of order $U^2 t^{3n_F}$; the lower graph is a contribution $t^{n_F}$.
}{0}
This chain graph is independent from $U$; 
regarding the lowest order contribution in $U$, there are no linear terms in $U$ as the interaction connect only fields with the same $x_2$: The lowest order contribution is given by, see Fig. 2
\bea
&&A(\pp_{x_2}^+)=\int d\kk_1 d\kk_2 G_{x_2-n_F,x_2}(\kk_1)\times\nn\\
&&G_{x_2,x_2-n_F}(\kk_2)  G_{x_2,x_2-n_F}(\kk_1+\kk_2-\pp_{x_2}^+)
\eea
with $\pp_{x_2}^+=(0,0,p_{x_2}^+)$. Similar contributions appears integrating out the lower scale. It is 
convenient to add and subtract a factor
\be
M=\sum_{x_2} \a_{x_2}\int d\xx
(\psi^+_{+,\xx,x_2-n_F}\psi^-_{-,\xx,x_2}
+\psi^+_{-,\xx,x_2}\psi^-_{+,\xx,x_2-n_F})
\ee
which is included in the free integration.
We include such term in the free integration, and we set 
$P(d\psi^{\le 0})e^M\equiv \tilde P(d\psi^{\le 0})$, with 
$\tilde P(d\psi^{\le 0})$ with propagator, if
$\o_1=-; \o_2=+$ and $\d_1=0, \d_2=-1$
\bea
&&<\psi^-_{\o_i,\kk',x_2+\d_i n_F}\psi^+_{\o_j,\kk',y_2+\d_j n_F}>=
\d_{x_2,y_2}\chi_0(\kk')\times\\ 
&&\begin{pmatrix}
&-i
k_0- v_F\sin k'+c(k') & \s_{x_2}\\ &\s_{x_2}
&-i k_0+v_F \sin k' +c(k')
\end{pmatrix}^{-1}_{i,j}\nn\label{prop1}
\eea
We consider $\s_{x_2}$ and $\a_{x_2}$ as independent, and we will choose  
$\a_{x_2}$ as function of $U$ and $\s$
so that the flow of the corresponding coupling is bounded; at the end we impose the condition
\be
\s_{x_2}=\a_{x_2}
\ee
We describe our RG analysis inductively.  We write $\psi^{(\le 0)}_\o=
\sum_{h=-\io}^0 \psi^h_\o$ and the corresponding propagator has cut-off $f_h$
with support in $\g^{h-1}\le |\kk'|\le \g^{h+1}$ with $\g>1$ a momentum scale.

After the integration of $\psi^{(0)},...\psi^{(h-1)}$ one gets that the generating function has the form
\be
\int  P(d\psi^{(\le h)}) e^{\VV^{(h)}(\psi^{\le h},\phi)}\label{eff}
\ee where the propagator is 
\bea
&&<\psi^-_{\o_i,\kk',x_2+\d_i n_F}\psi^+_{\o_j,\kk',y_2+\d_j n_F}>={\d_{x_2,y_2}\over 
Z_1^{(h)}}
\chi_h(\kk')\\ 
&&\begin{pmatrix}
&-i
k_0- v_{h}\sin k'+c(k') & \s_{x_2}\\ &\s_{x_2}
&-i k_0+v_{h} \sin k' +c(k')
\end{pmatrix}^{-1}_{i,\j}\nn\label{prop1}
\eea
and $\VV^{(h)}(\psi,0)=$
\be \sum_{m,\underline\o} \sum_{x_{2,1},.., x_{2,m}}
\int d\kk'_1...d\kk'_m  W_{m}^{(h)}(\underline \kk')\prod_i \psi^{\e_i(\le h)}_{\o_i,\kk'_i,\xx_{2,i}}\d_{m}
\label{ep}\ee
where $\d_m$ vanishing in correspondence of \pref{delta}; $Z_h$ is a wave function renormalization, $v_h$ is an effective Fermi velocity and $\chi_h=
\sum_{k\le h} f_k$  with support in $|\kk'|\le \g^{h+1}$; $\VV^{(h)}(\psi,\phi)$ as a similar expression as \pref{ep} with some 
of the fields $\psi$ replaced by external fields $\phi$.

We have to extract from the effective potential the relevant and marginal terms, which contribute to the corresponding running coupling constants.
The scaling dimension of the theory is $D=2-n/2$, so all the terms with $n\ge 6$ are irrelevant. If we renormalize all the 
quartic terms $\psi^+_{\o_1,x_{2,1}}\psi^-_{\o_2,x_{2,2}}
\psi^+_{\o_2,x_{2,3}}\psi^-_{\o_3,x_{2,4}}$ we would get a huge number of running coupling constants, one for any choice of
$\o_1,..,\o_4$ and $x_{2,1},..,x_{2,4}$. There is however a dramatic improving with respect to power counting, and a huge class of quadratic or quartic terms are indeed irrelevant, namely:
\begin{enumerate}
\item The terms such that the r.h.s. of \pref{delta} is non vanishing;
\item The quartic terms with different $x_{2,i}$, and the marginal
quadratic terms  with different $x_{2,i}$.
\end{enumerate}
Condition (1) is quite 
natural in the commensurate case $\a=p/q$; indeed if it is violated than the corresponding process disappear at scales smaller that some energy scale $\bar h=O(\log 1/q)$ by conservation of momenta measured from the Fermi points. In the incommensurate case
things are however more subtle. The l.h.s. of \pref{delta}
can be arbitrarily small and there is no a finite scale below which such terms disappear. In other terms, there are quadratic processes which connect with arbitrary precision Fermi points  
$p_{\o}^{x_2}$ can be arbitrarily close to $p_{\o'}^{x'_2}$  for large $x_2-x'_2$; deciding if they are relevant or irrelevant is a rather subtle issue which will be discussed below, and it can depend on the specific form of the considered quasi periodic system. Condition 2), on the other hand, depends on the presence of a gap.

We introduce a renormalization operation which acts on the quadratic or quartic terms.
Regarding the quadratic terms, condition (1) says the non irrelevant terms verify
\be (\o_1-\o_2)p_F+2\pi\a (x_{2,1}-x_{2,2})=0\ee If $\o_1=\o_2$
we define a renormalization operation $\RR$ consisting in extracting
from the kernel $W^h(\kk)$ the term 
$W^h(\pp_{\o}^{x_2})+(k-p_{\o}^{x_2})\partial W^h (\pp_{\o}^{x_2}) +k_0\partial W^h(0)$. The first term contributes to the renormalization of the chemical potential
\be
F_\n^{(h)}=\sum_{\o,\s}\sum_{x_2} \int d\xx \g^h \n_{x_2}\psi^+_{\underline x,\o,\s}
\psi^-_{\underline x,\o,\s}
\ee
while the other terms contribute to the wave function, that is $Z_{h-1}=Z_h(1+\partial_0 W^h)$, and Fermi velocity renormalization.

On the other hand if $\o_1=-\o_2=\pm$ the r.h.s. of 
\pref{delta} is vanishing if $n_F=(x_{2,2}-x_{2,1})$ and
$p_{-}^{x_2}=p_{+}^{x_2-n_F}$;
we define the renormalization operation $\RR$ in this case
as the subtraction from the kernel $W^h(\kk)$ of the term 
$W^h(\pp_{\o}^{x_2})$ and this produces an effective interaction
\be 
F_\a^{(h)}=\int d\xx 2^h \a_{x_2}(
\psi^+_{+,x_2-n_F}\psi^-_{-,x_2}
+\psi^+_{-,x_2}\psi^-_{+,x_2-n_F})\ee 
Regarding the quartic terms, the $\RR$ operation is non trivial only 
on the quartic terms with the same $x_2$, and in such a case 
we extract from $W_4^h(\kk_1,\kk_2,\kk_2)$
the term $W_4^h(\pp_{\o_1}^{x_2},\pp_{\o_2}^{x_2},
\pp_{\o_3}^{x_2},\pp_{\o4}^{x_2})
$.  The effective potential can be therefore written as
$\VV^{(h)}=\LL \VV^{(h)}+\RR \VV^{(h)}$ where
$\LL \VV^{(h)}$ is the relevant or marginal part 
\be
\VV^{(h)}(\psi,0)=F_\n^{(h)}+F_\a^{(h)}+F_1^{(h)}+F_2^{(h)}+F_4^{(h)}
\ee
with 
\bea
&&F_1^{(h)}=\sum_{x_2,\s,\s',\o}\int d\xx
g_{1,h,x_2}
\psi^+_{\underline x,\o,\s}\psi^-_{\underline ,-\o,\s}\psi^+_{\underline x,-\o,\s'} \psi^-_{\underline x,\o,\s'}\nn\\
&&F_2^{(h)}=\sum_{x_2,\s,\s',\o}\int d\xx
g_{2,h,x_2}\psi^+_{\underline x,\o,\s}\psi^-_{\underline x,\o,\s}\psi^+_{\underline x,-\o,\s'}\psi^-_{\underline x,-\o,\s'}\nn\\
&&F_4^{(h)}=\sum_{x_2,\s,\s',\o}\int d\xx
g_{4,h,x_2}
\psi^+_{\underline x,\o,\s}\psi^-_{\underline x,\o,\s}
\psi^+_{\underline x,\o,\s'}\psi^-_{\underline x,\o,\s'}\nn
\eea
Note that the quartic marginal terms in $\LL V^h$ only
connect fermions with the same $x_2$, that is in the same wire;
all the processes connecting different wires are irrelevant.
The only terms connecting different wires are the hopping terms.
Integrating the field $\psi^h$ one gets an expression 
similar to \pref{eff} with $h$ replaced by $h-1$
and the procedure can be iterated.

We have to discuss the flow of the running coupling constants. Note that the RG flow stops at a scale $h^*=-\log \s$.
One has first to fix the counterterms $\a,\n$ so that the flow of the relevant running coupling constants is bounded.
We write
\be
\a_{h-1}=\g\a_h+\b^h_\a
\ee
where in $\b^h_{\a}$ one can separate two kinds of terms: a)
the ones independent from $U$, which are
$O(t^{n_F}\g^{\th k})$ (the factor $\g^{\th k}$, $0<\th<1$ follows from the irrelevance 
of the $t$ vertices, see the following section); b) the ones
with at least one $U$ or $g_{i,k}$ quartic coupling, which are at least 
quadratic in $U$ (both the initial interaction $V$ and the quartic effective interactions
in $\LL V^k$ involve fields with the same $x_2$) and  $O(U^2 \s^{3}\g^{-3 h})$.
Therefore we can choose
$\a_0$ so that the flow is bounded, that is 
$\a_0=-\sum_{k=h^*}^0 \g^k \b^k_{\a}$
and the r.h.s.
is bounded by
$\sum_{k=h^*}^0 (\g^k t^{n_F}\g^{\th k}+
U^2 \s^{3}\g^{-2 k})$ and finally, extracting the dominant term 
\be
\a_0=t^{n_F}(a_{n_F}+R) \quad\quad |R|\le  C(t+U^2)
\ee
 and $t^{n_F}a_{n_F}$
is the contribution from the chain graph, see Fig. 2
\be
a_{n_F}=\prod_{k=1}^{n_F-1}{1\over \cos(-n_F\pi\a+2\pi \a k)-\cos(n_F\pi \a)}
\ee
which is independent from $x_2$; moreover $\a_h$ behave as
$t^{n_F}\g^{\th h}+
U^2 \s^3 \g^{-3 h}$.
%

Similarly we have to control the flow of $\n_h$; we write
$\n_h=\g \n_{h+1}+\b_\n^h$ with $\b_h^\n$ is sum of terms
$O(U \g^{\th h})$ (the contributions independent on $t$, where the 
$\g^{\th h}$ comes from a parity cancellation) and $O(t\g^{\th h})$ (the terms containing $t$ vertices) or $O(U \s^2 \g^{-2 h})$; in order to have $\n_h$ small we choose a $\n_0$ so that $\n_0=-\sum_{k=h^*}^0 \g^k \b_k$ and 
$|\n_0|\le C (U+t)$ and $\n_h$ behave as $t\g^{\th h}+U \s^2 \g^{-2 h}$.

In order to discuss the flow of the quartic running coupling constants $g_{1,h},g_{2,h},g_{4,h}$,
we notice that we can write $g_{i,h-1}=g_{i,h}+\b_{i,1}^h+\b_{i,2}^h$ with 
$\b_{i,1}^h$ sum of graphs containing only quartic vertices $g_{1,k}$
and $\b_{i,2}^h$ with at least a vertex $t,\n_k,\a_k,\s$. By iteration, if $i=2$ 
$g_{i,h-1}=g_{i,0}+\sum_{k=0}^h(\b_{2,1}^k+\b_{2,2}^k)$ and
the second addend is bounded by 
$\sum_{k=0}^h U^2 (\a_h+\n_h)$ hence is $O(U^2)$ while $\b_{2,1}^h$ again is summable as 
is proportional to $g_{1,h}^2$; therefore $g_{2,h}, g_{4,h-1}$ tends to values which are $U+O(U^2)$.
On the other hand 
$g_{1,h}\sim {U\over 1-a U h}$, that is tends to vanish for repulsive interactions while
$v_h\to v_{-\io}=v_F(1+O(U))$; finally the wave function renormalization behaves as $Z_h\sim \g^{\h h}$ with
$\h=b U^2+O(U^2)$,$b>0$. By imposing the condition $\a=\s$ one gets the size of the gap in the interacting case.

It is finally convenient to compare the above flow with the in one dimensional models. In the interacting Aubry-Andre' model the flow of the gap term is linear in the effective coupling, as the quasi-periodic potential involve fermions on the same chain; therefore one has a contribution to the analogous of $\b_\a$ of the form
$\s \g^{-k}$ which corresponds to the generation of anomalous critical exponents in the gaps  \cite{M1}. In interacting fermionc Fibonacci chains one considers infinitely many
quadratic couplings and this produce a complex flow suggest the closure  of all gaps except a finite number in the attractive case  \cite{G1},\cite{G2}.

\section{Convergence}

As we discussed before the presence of small divisors in the expansions
has the effect that information on persistence of gaps are encoded in the convergence or divergence of the whole renormalized series; in particular, one has to 
discuss the relevance or irrelevance that the Umklapp terms almost connecting Fermi
points.
 
The kernels of the effective potential $V^h$ can be written as sum of graphs
such that to each line connecting two points $\underline x$ with $\underline y$ is associated a scale $h$ and it corresponds to a propagator $\d_{x_2,y_2}  
\bar g^{(h)}(\xx,\yy)$ defined by \pref{prop1}; to the vertices are associated 
the effective couplings $g_h,\d_h,\n_h,\a_h$ and the couplings $\l, t,\n$.
The scales induce a structure of {\it clusters} in the graph;
each cluster $v$ with scale $h_v$ contains a connected 
subset of the graph, such that the internal propagators have scale 
$\le h_v$ and at least one of them scale $h_v$, and the external lines scales $>h_v$; the clusters can be represented as a tree $\t$, see Fig.3.
We call $S_v$ the number of subclusters $w$ in the cluster $v$, with $w'=v$, connected by $S_v-1$ propagators $g^{(h_v)}$.
We associate a scale $h_v$ also to the end-points and $v'$ 
is the first cluster enclosing it;
 regarding the end-point $g_h,\d_h,\n_h$ one has $h_{v'}=h_v-1$.
We call $\bar m_v^i$, $i=t,g_h,\n_h,\a_h$
the number of $i$ end-points in$v$ and not 
not contained in other smaller clusters,
and $m_v^i$, $\a=t,g_h,\n_h,\a_h$ the total number of $i$ end-points in $v$. To each cluster $v$ is associated a set of $p_v$ external lines with 
scale $<h_v$ and coordinate $\xx_i$.

\insertplot{800}{200}
{}
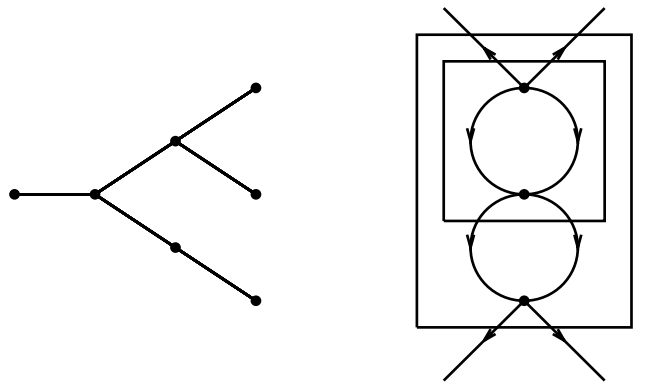
{fig9a}{\label{n11} A graph with its clusters and the corresponding tree
}{0}

We can define two kind of clusters:
\begin{enumerate}
\item
The {\it non resonant} clusters  $v\in N R$ are such that $\sum_i 
\e_i p^{\o_i}_{x_{2,i}}\not =0$
\item
The {\it resonant} clusters  $v\in R$ are such 
that $\sum \e_i p^{\o_i}_{x_{2,i}}=0$; 
$v\in R1$ are such that all the $x_{2,i}$ of the external lines are equal;
$v\in R2$ are such that all the $x_{2,i}$ of the external lines are not all equal.
\end{enumerate}
According to the previous definitions, the $\RR$ operation acts non trivially
only on the clusters $v\in R1$ with $2$ or $4$ external lines
or $v\in R2$ with two external lines. In the quartic terms
the action of $\RR$ consists in replacing an external field
$\psi_\xx$ with $\psi_\xx-\psi_\yy=(\xx-\yy)\int_0^1 dt \partial \psi$;
the same action is for the terms with two external lines $v\in R2$, while
there is a replacement with the second difference when $v\in R1$ and two external lines.
With respect to the $\RR=0$ case, this corresponds to an extra derivative on the external lines, giving a factor $\g^{h_{v'}}$ and an extra 
$(\xx-\yy)$ which can be associated to the propagators $g^{h_{v}}$
and produces dimensionally a factor $\g^{-h_{v}}$. 
The same factor is obtained in quadratic terns $v\in R2$
while in the quadratic term $v\in R1$ the second difference produces a term
$\g^{2(h_{v'}-h_v)}$.
In conclusion
the $\RR$ operation produces a factor  $\g^{z_v(h_{v'}-h_v)}$
with a) $z_v=1$ if $p_v=4$ $v\in R1$; b)
$z_v=2$ if $p_v=2$ and $v\in R1$; b)
$z_v=1$ if $v\in R2$ and $p_v=2$; $z=0$ in all the other cases.

\insertplot{520}{130}
{
\ins{100pt}{40pt}{$v'$}
\ins{175pt}{40pt}{$v$}
}
{figjsp467}
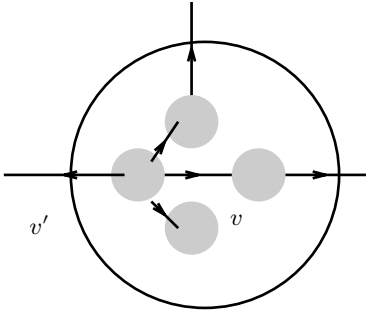
{\label{h2}  A representation of a cluster $v$ and the $S_v$ subclusters.
The lines internal to the blob have scale $h_v$, the lines external $h_ {v'}$; the gray blobs have a similar structure and so on.
} {0}

The size of a generic Feynman graph is easily obtained using that $|g^h(\xx)|\le C\g^{h}$
and $\int d\xx |g^h(\xx)|\le C\g^{-h}$; 
by choosing in the graph a tree of propagators connecting the $S_v$
clusters or end-points, see Fig. 4,  we get by integrating a factor $\g^{-h_v (S_v-1)}$ while the remaining propagators are bounded by 
$\g^{h_v (n_v-S_v+1)}$, where $n_v$ is the number of propagators $g^{h_v}$: note that the sum over $x_2$ is done using the
kronecker deltas in the propagator of the tree, causing that only one sum remain. The bound for the Feynman graph is proportional to, up to
a constant $C^m$, $m$ is the number of vertices and not taking into account the $\RR$ operation
\bea
&&
\prod_v \g^{-2 h_v (S_v-1)} \prod_v \g^{n_v h_v}
\prod_{v} (\n_{h_v} \g^{h_v})^{\bar m_v^\n}\nn\\
&&\prod_{v} t^{\bar m_v^t}\prod_{v} (\a_{h_v} \g^{h_v})^{\bar m_v^\a}
=\g^{ (2-n/2)h}
\prod_v \g^{ -(h_v-h_{v'})D_v}\nn\\
&&\prod_{v} (t \g^{-h_v})^{\bar m_v^t}  
\prod_{v} (\n_{h_v})^{\bar m_v^\n}\prod_{v} (\a_{h_v})^{\bar m_v^\a}\label{rr}
\eea
with $D_v=2-n^e_v/2$ and $n^e_v$ is the number of 
external lines of $v$. 
In principle a bound on Feynman graphs is not enough for getting
non-perturbative information; even if a finite bound is obtained at order $,$,
one has to worry about extra combinatorial $m!$ due to the large number of graphs
which could ruin convergence. It is however a well known fact that cancellations
due to Pauli principle in fermionic expansions has the effect that such extra $m!$
are absent.
We get therefore the following estimate,  if $\e=max(|U|, t^{1\over 2})$
and using that the $g_{i,h}$ are bounded by bare coupling $U$ times a constant, if $U>0$, as discussed in the previous section
\bea
&&{1\over L\b}\int d\xx |W^h(\xx)|\le \sum_m \e^m\sum_{\t, h_v,n_v}
\g^{ (2-n/2)h}\nn\\
&&[\prod_{v} (\s \g^{-h_v})^{\bar m_v^\s}][\prod_v \g^{ -(h_v-h_{v'})(D_v+z_v)}\prod_{v} (t^{1\over 2}\g^{-h_v})^{\bar m_v^t}  
\nn
\eea

where  we take into account
the effect of the $\RR$ 
operation and of the presence of non-diagonal propagators, 
giving extra factors $\prod_{v} (\s \g^{-h_v})^{\bar m_v^\s}$.
One needs to sum over all the possible attributions of scales $h_v$; the sum would be finite of $D_v+z_v$ can be vanishing or negative,
what however is not the case. This lack of convergence is a manifestation of the small divisor problem, as it is due also to the fact that we have not renormalized
the quadratic and quartic non resonant terms. In order to show that they give
a finite contribution one has to improve the estimate by
the Diophantine property of $\a$ \pref{d}. Let us consider a non resonant cluster $v\in NR$ with 2 external lines; we get, $\d=0,1$ 
\bea
&&2  \g^{h_{v'}}\ge ||k'_1||+||k'_2||\ge ||k'_1-k'_2||\nn\\
&&\ge ||2\d n_F \pi \a +2\pi \a(x_{2}-x'_{2})||\ge C_0 |x_2-x'_2|^{-\t}\nn
\eea
so that 
\be
|x_2-x'_2|\ge C \g^{-h_{v'}\over \t} 
\ee
This says that in order to have a cluster a low scales the difference of coordinates must be large. In addition, if we apply this to the $t$ vertices
when $x_2-x'_2=\pm 1$ it says that $h_v'$ is bounded by a constant so that \be \prod_{v} (t^{1\over 2} \g^{-h_v})^{\bar m_v^t}\le \prod_{v} (t^{1\over 2} C)^{\bar m_v^t}\ee 
Regarding the terms with 4 lines we can write
\bea
&& 4\g^{h_{v'}}\ge  ||\sum_i \e_i \kk'_i||\ge
||2\pi \a \sum_{i=1}^4 \e_i x_{2,i}+ \sum_i \e_i\o_i \pi n_F \a||\ge \nn\\
&&C_0 | \sum_{i=1}^4 \e_i x_{2,i}+\sum_i \e_i\o_i n_F |^{-\t}
\ge C |\bar x_2-\bar x'_2|^{-\t}
\nn
\eea
where $|\bar x_2-\bar x'_2|$ is the maximal difference of the $x_2$ of the incoming and outcoming lines; therefore 
\be
|\bar x_2-\bar x'_2|\ge C \g^{-h_{v'}\over \t}
\ee
Note that there is a path of propagators connecting the external lines with coordinates $\bar x_2$ and $\bar x'_2$ and \be 
|\bar x_2-\bar x'_2|\le n_F N_v+m^t_v\le 2n_F N_v
\ee  
where $N_v$ is the number of vertices in the cluster $v$; 
the reason is that one modify the coordinate by non diagonal
propagators or vertices $t$. In conclusion
\be
N_v\ge C \g^{-h_{v'}\over \t}/n_F^{1\over \t}\label{jj}
\ee
We can now associate to each vertex in the graph a constant $\bar c<1$ (at the 
expense of a factor $\bar c^{-m}$ in the final bound). Moreover we can write 
$\bar c
=\prod_{h=-\io}^1 \bar c^{2^h/2}$ so that we can associate a factor 
$c^{2^h_n/2}$ to each of the $N_v$ vertices contained in a cluster $v$; therefore
\be
\bar c^m\le \prod_v \bar c^{N_v 2^{h_v}}\le \prod_v \bar c^{N_v 2^{h_{v'}}}
\ee
and using \pref{jj} one gets
\be
\bar c^m\le \prod_{v\in NR} \bar c^{ C\g^{-h_{v'}\over \t }2^{h_{v'}}/n_F^{1\over \t}}\le \bar C^n \prod_{v\in NR}  \g^{2(h_{v'}-h_v)}
\ee
provided that $\g^{1\over \t}/2 =\g^{\bar \x}.$ with $\bar \x>0$ ($\g>1,\t>1$), and we have used $e^{-\a x} x^N\le
(N e/\a)^N$ with $x=\g^{-\bar \x h}$. We can choose for instance 
$\g^{1\over \t}=4$, $\g^{\bar \x}=2$.

We have finally to consider the quartic terms or the marginal quadratic terms $v\in R2$. We note first that due to the presence of a gap there is a scale $h^*=-\log \s$,
with $\s=O(t^{n_F})$, such that the fields $\le h^*$ can be integrated in a single step; that is, the iterative integration stops at $h^*$.
As the external lines of the clusters $v\in R2$ have different coordinate $x_2$,
necessarily contain a non diagonal propagator or a $t$ or $\a$ end-point;
in the first case one of the factors \pref{rr}
$(\s \g^{-h_v})\le \g^{(h^*-h_v)}$ provides the dimensional gain of all the clusters containing such non diagonal propagator. If there is a $t$ vertex
we use $t^{1\over 2}\le \g^{(h^*-h_v)\over 2 n_F}$. Similarly is there is an $\a$ vertex we use that $\a_h$ is $O(\s^2 U\g^{-h})$ or $O(t U\g^{\th h})$
one gets an extra  $\g^{(h^*-h_v)\over 2 n_F}$.

In conclusion
\bea
&&{1\over L\b}\int d\xx |W^h(\xx)|\le\\
&&\sum_m \sum_{\t, h_v,n_v}
\g^{ (2-n/2)h}\e^{m} [\prod_v \g^{ -(h_v-h_{v'})(D_v+\bar z_v)}
\nn
\eea
where
\begin{itemize}
\item $\bar z_v=2$ if $v\in NR$ and $n^v_e=4,2$
\item $\bar z_v=1$ if $v\in R1$ and $n^v_e=4$, $z_v=2$
if $v\in R1$ and $n^v_e=2$
\item $z_v=1+1/n_F$
if $v\in R2$ and $n^v_e=2$; $z_v=1/n_F$
if $v\in R2$ and $n^v_e=4$ .
\end{itemize}
Therefore we can sum over the scales and one gets a convergent estimate for the effective potential; moreover the contributions with an irrelevant $t$ coupling have an extra $\g^{\th h}$ due to the fact that the dimensions are all negative.

It is immediate to get the large distance asymptotic decay of the 2-point function. 
The decay in $\xx$ is an immediate consequence of the fact that there is a last scale $h^*$; the decay rate $\s$ provide an estimate in the gap
of the interacting case, which is always non vanishing for $U$ small.
The decay in the direction $x_2$  is faster than any power with rate $\log t$ because
the contribution in $t$ starts from order $x_2-x'_2$.

\section{Conclusions}

We have proven the persistence of the gaps in the interacting anisotropic Hofstadter model with small hopping $t$, that is close to the multi-wire
limit. When the hopping is small the non-interacting gaps are estimated by their lowest order contribution $a_{n_F} t^{n_F}$, and they persist in presence of interactions even when the coupling $U$ is much smaller
than the non interacting gaps $U^2>>a_{n_F} t^{n_F}$. 
The main difficulty relies in the 
presence of infinitely many processes which, due to Umklapp scattering and the incommensurability of the two periods, connect arbitrarily close the 
Fermi points. We can however rigorously establish the 
irrelevance of such terms by combining non perturbative RG
methods with a strategy inspired by KAM problems and relying on number theoretical properties of irrationals.  
This has the effect that there is only a small number of quadratic running coupling constants, in contrast to the one dimensional Fibonacci chains \cite{G1}, \cite{G2}.
Therefore the validity of the mechanism proposed in \cite{G1}, \cite{G2} for the generation of a finite measure spectrum is excluded, but on the other hand we can prove the persistence of gaps
only for $U$ smaller than $a_{n_F}$ and not uniformly.
An important open problem is what happens to the gap in the case of larger hopping $t$, when the lowest order dominance is not true; even in the non interacting case this a quite hard non-perturbative issue requiring different methods \cite{A}. Other interesting issues include 
what happens to gaps in
the case of attractive potential $U<0$ or when $U^2>>a_{n_F}$. One could consider also the case of chemical potentials in the spectrum of the non interacting case, and
investigate the question of the generation of gaps due to the interaction.
The same argument explained above shows that the non resonant terms terms are irrelevant, but resonant terms connecting different wires are instead marginal
and have a complicate flow which could exhibit non trivial fixed points.
This opens the way to the a quantitative understanding 
starting from a microscopic lattice model
of the opening of new gaps caused by the interaction, as it appears in experiments
\cite{1}-\cite{4}.

\bibliographystyle{amsalpha}

\end{document}